\documentclass[aps,10pt,showpacs,prd,superscriptaddress,twocolumn,nofootinbib]{revtex4-1}

\usepackage{graphicx}	
\usepackage{amsmath}	
\usepackage{amssymb}	
\usepackage{xcolor}
\usepackage{soul}
\definecolor{linkcolor}{rgb}{0.0,0.3,0.5}
\usepackage[unicode,colorlinks=true,citecolor=linkcolor,linkcolor=linkcolor,urlcolor=linkcolor]{hyperref}

\newcommand{\rmi}{\mathrm{i}} 
\newcommand{\e}{\mathrm{e}} 

\begin{document}

\title{Long-Lived Inverse Chirp Signals from Core
Collapse in Massive Scalar-Tensor Gravity}

\author{Ulrich Sperhake}
\email{us248@maths.cam.ac.uk}
\affiliation{DAMTP, Centre for Mathematical Sciences, University of Cambridge, Wilberforce Road, Cambridge CB3 0WA, United Kingdom}
\affiliation{{TAPIR 350-17, Caltech, 1200 East California Boulevard, Pasadena, California 91125, USA}}

\author{Christopher J. Moore}
\affiliation{DAMTP, Centre for Mathematical Sciences, University of Cambridge, Wilberforce Road, Cambridge CB3 0WA, UK}
\affiliation{IST-CENTRA, Departamento de F{\'i}sica, Avenida Rovisco Pais 1, 1049 Lisboa, Portugal}

\author{Roxana Rosca}
\affiliation{DAMTP, Centre for Mathematical Sciences, University of Cambridge, Wilberforce Road, Cambridge CB3 0WA, UK}

\author{Michalis Agathos}
\affiliation{DAMTP, Centre for Mathematical Sciences, University of Cambridge, Wilberforce Road, Cambridge CB3 0WA, UK}

\author{Davide Gerosa}
\thanks{Einstein Fellow}
\affiliation{{TAPIR 350-17, Caltech, 1200 E. California Boulevard, Pasadena, California 91125, USA}}
\thanks{Einstein Fellow}

\author{Christian D. Ott }
\affiliation{{TAPIR 350-17, Caltech, 1200 E. California Boulevard, Pasadena, California 91125, USA}}

\date{\today}

\begin{abstract}
This Letter considers stellar core collapse in massive scalar-tensor theories of gravity. The presence of a mass term for the scalar field allows for dramatic increases in the radiated gravitational wave signal. There are several potential \emph{smoking gun} signatures of a departure from general relativity associated with this process. These signatures could show up within existing LIGO-Virgo searches.
\end{abstract}

\maketitle 

\noindent{\bf \em Introduction~--~}
General relativity (GR) has successfully passed numerous tests \cite{Psaltis:2008bb,Will:2014kxa}
and, in the words of Ref.~\cite{Alsing:2011er},
``occupies a well-earned place next to the standard model as
one of the two pillars of modern physics.''
And yet, the enigmatic nature of {\em dark energy} and {\em dark matter}
evoked in the explanation of cosmological and astrophysical observations
\cite{Spergel:2015noa}, as well as theoretical considerations
regarding the renormalization of the theory in a quantum theory sense,
indicate that GR may ultimately need modifications in the low- and/or
high-energy regime \cite{Berti:2015itd}.

Tests of GR have so far been almost exclusively limited to relatively weak fields.
But the recent breakthrough detection of gravitational
waves (GWs) by LIGO \cite{Abbott:2016blz} has opened a new
observational channel towards strong-field gravity, and tests
of Einstein's theory are a key goal of the new field of GW physics
\cite{TheLIGOScientific:2016src,Yunes:2016jcc}. 
Most GW-based tests either (i) construct a phenomenological parameterization of possible deviations from the expected physics and seek to constrain the different parameters,
or (ii) model the physical system in the framework of a chosen alternative theory to see if it can better explain the observed data.

The latter approach faces significant challenges;
the candidate theory must agree with
GR in the well-tested weak-field regime and yet lead to measurable
strong-gravity effects. Furthermore a mathematical
understanding of the theory, in particular its well-posedness,
is necessary for fully nonlinear simulations.
One of the most popular candidate extensions of GR are
scalar tensor (ST) theories of gravity
\cite{Damour:1992we,Fujii:2003pa},
adding a scalar sector to the vector and tensor fields 
of Maxwell GR. Scalar fields naturally arise in
higher-dimensional theories including string theory and feature prominently in cosmology, and ST theories have a well-posed Cauchy formulation.
ST theories also give rise to the most concrete example of a
strong deviation from GR known to date:
the {\em spontaneous scalarization} of neutron stars
\cite{Damour:1993hw}.
The magnitude of this effect facilitates strong constraints on the parameter space of ST theory through binary pulsar observations \cite{Freire:2012mg,Antoniadis:2013pzd,Wex:2014nva}.
These bounds, as well as the impressive
constraints obtained from the Cassini mission
\cite{2003Natur.425..374B}, however, are all based on
observations of widely separated objects and, therefore, apply
only to massless ST theory [or theories
with a scalar mass $\mu\!\lesssim\!10^{-19}\,{\rm eV}$ 
yielding a Compton wavelength, $\lambda_{\textrm{c}}\!=\!(2\pi \hbar)/(\mu c)$, greater than or comparable to the objects' separation \cite{Alsing:2011er,Ramazanoglu:2016kul}].

Deviations of black-hole spacetimes from GR
are limited in ST
gravity due to the \textit{no-hair} theorems
\cite{Hawking:1972qk,Thorne:1971},
although we note that scalar radiation has been observed in
black-hole binary simulations
for nontrivial scalar potentials
\cite{Healy:2011ef} or boundary conditions \cite{Berti:2013gfa}.
Nevertheless, the most straightforward way to bypass the no-hair
theorems is to depart from vacuum.
Neutron stars and stellar core collapse thus appear to be
the most promising systems to search for characteristic signatures;
cf.~\cite{Novak:1999jg,Palenzuela:2013hsa,2016CQGra..33m5002G} and references therein.

Here, we perform the first study of dynamic strong-field
systems in {\em massive} ST theory through exploring
GW generation in core collapse.
As we will see below, the GW signal is dominated by the
rapid phase transition from weak to strong scalarization
and the ensuing dispersion of the signal.
We therefore focus in this study on spherically symmetric models
which capture the key features of the collapse responsible for
spontaneous scalarization.

The most promising
range of the scalar field mass $\mu$ for
generating strong scalarization and satisfying existing
binary pulsar constraints has been identified
as $\mu \! \gtrsim \! 10^{-15}\,{\rm eV}$
\cite{Ramazanoglu:2016kul,Morisaki:2017nit}.
In massive ST theory, low-frequency modes with $f< f_*=\mu/(2\pi \hbar)$
decay exponentially with distance rather than radiate towards infinity.
For masses
$\mu>10^{-13}\,{\rm eV}$
($f_*>24.2\,{\rm Hz}$), the
GW power detectable inside the
LIGO sensitivity window $10\,{\rm Hz}\!\lesssim\! f
\lesssim\! 10^3\,{\rm Hz}$ would be considerably reduced due to this effect.
We therefore study in this work
the range
$10^{-15}\,{\rm eV} \lesssim \mu \lesssim 10^{-13}\,{\rm eV}$.

\vspace{0.12cm}
%
%
\noindent{\bf \em Formalism~--~}
%
The starting point of our formulation is the generic action for a scalar-tensor
theory of gravity that (i) involves a single scalar field nonminimally coupled to the metric, (ii) obeys the covariance principle,
(iii) results in field equations of at most second differential order, and (iv) satisfies the weak equivalence principle. In the Einstein frame,
the action can be written in the form (using natural units $G=c=1$) \cite{Berti:2015itd,Fujii:2003pa}
\begin{eqnarray}
  \!\!S&=&\!\!\int \!\!dx^4\frac{\sqrt{-\bar{g}}}{16\pi} [\bar{R}
    -2\bar{g}^{\mu\nu}\partial_{\mu}\varphi\,\partial_{\nu}\varphi
    - 4V(\varphi)] + S_m,
\end{eqnarray}
where $\varphi$ is the
scalar field,
$V(\varphi)$ the potential, and $\bar{R}$ and $\bar{g}$
the Ricci scalar and determinant
constructed from the conformal metric $\bar{g}_{\mu\nu}$, respectively.
$S_m$ denotes the contribution due to matter fields,
that couple to the physical or
Jordan-Fierz metric ${g_{\mu\nu}\!=\!\bar{g}_{\mu\nu}/F(\varphi)}$,
$F(\varphi)$ the coupling function,
and the
physical energy momentum tensor is ${T^{\mu\nu}\!=\!2(-g)^{-1/2}
\delta S_m/\delta g_{\mu\nu}}$, assumed here to describe a perfect fluid
with baryon density $\rho$, pressure $P$, internal energy $\epsilon$,
enthalpy $H$, and ${\textrm{4-velocity}}$ $u^{\alpha}$:
\begin{equation}
  T_{\alpha\beta} = \rho H u_{\alpha}u_{\beta} + Pg_{\alpha\beta}\,,
  ~~~~~
  H = 1+\epsilon + P/\rho\,.
\end{equation}
The equations of motion are given by
\begin{eqnarray}
  &&\bar{G}_{\alpha\beta} = 2\partial_{\alpha}\varphi\,
    \partial_{\beta}\varphi-\bar{g}_{\alpha\beta} \partial^{\mu}
    \varphi \,\partial_{\mu}\varphi + 8\pi \bar{T}_{\alpha\beta}
    -2V\bar{g}_{\alpha\beta}\,, \nonumber \\
  && \bar{\nabla}^{\mu}\bar{\nabla}_{\mu} \varphi =
    2\pi (F_{,\varphi}/F) \bar{T}+V_{,\varphi}\,, \nonumber\\
  && \bar{\nabla}_{\mu}\bar{T}^{\mu\alpha}=
    -\frac{1}{2} \frac{F_{,\varphi}}{F} \bar{T}\,\bar{g}^{\alpha\mu}
    \partial_{\mu} \varphi\,,~~~~~\nabla_{\mu}(\rho u^{\mu})=0\,,
\end{eqnarray}
where the conformal energy momentum tensor is $\bar{T}_{\alpha\beta}
=T_{\alpha\beta}/F$, $\bar{\nabla}$ is the covariant derivative
constructed from $\bar{g}_{\mu\nu}$, the subscript
${,\varphi}$ denotes $d/d\varphi$ and the
last equation arises from conservation of the matter current density in
the physical frame.

Henceforth, we assume spherical symmetry, writing
\begin{equation}
  d\bar{s}^2=\bar{g}_{\mu\nu}dx^{\mu}dx^{\nu}=-F\alpha^2 dt^2
    +FX^2 dr^2 + r^2 d\Omega^2\,,
\end{equation}
where $\alpha\!=\!\alpha(t,r)$, $X\!=\!X(t,r)$ and we also define for
convenience $\Phi\!=\!\ln (\sqrt{F}\,\alpha)$ and the gravitational
mass $m\!=\!r[1-(FX^2)^{-1}]/2$.
In spherical symmetry, the ${\textrm{4-velocity}}$ in the Jordan frame is
  \mbox{$u^{\mu}=(1-v^2)^{-1/2}~[\alpha^{-1},~v\,X^{-1},~0,~0]\,,$}
where the velocity field $v$ as well as the other matter variables
$\rho$, $P$, $\epsilon$ and $H$ are also functions of $(t,r)$.
High-resolution shock capturing requires a flux conservative formulation of the matter equations which is achieved by (cf.~\cite{2016CQGra..33m5002G}) changing from variables
$(\rho,\,v,\,H)$ to
\begin{equation}
  D \! = \! \frac{\rho X F^{-3/2}}{\sqrt{1-v^2}}\,,\;
  S^r \! = \! \frac{\rho H v F^{-2}}{(1-v^2)}\,,\;
  \tau \! = \! \frac{S^r}{v}-\frac{P}{F^2}-D\,.
\end{equation}
Finally, we introduce $\eta=X^{-1}\,\partial_r \varphi$ and
$\psi=\alpha^{-1}\,\partial_t \varphi$. The resulting system of equations
is identical to Eqs.~(2.21), (2.22), (2.27), (2.28), and (2.33)-(2.39)
in Ref.~\cite{2016CQGra..33m5002G} except for the following additional
potential terms (bracketed numbers
denote right-hand sides in Ref.~\cite{2016CQGra..33m5002G}):
\begin{eqnarray}
  \partial_r \Phi &=& [2.21] -rFX^2V\,, \nonumber \\
  \partial_r m &=& [2.22] + r^2 V \,, \nonumber \\
  \partial_t \psi &=& [2.28] - \alpha F V_{,\varphi}\,,\nonumber \\
  s_{S^r} &=& [2.38] - rV\alpha X F\left(S^r v-\tau-D+F^{-2}P\right)
    \,,
\end{eqnarray}
where $s_{S^r}$ is the source term in the evolution of $S^r$.
All other equations in the above list remain unaltered.

We have implemented these equations by adding the
potential terms to the {\sc gr1d} code originally
developed in Ref.~\cite{O'Connor:2009vw} and extended
to massless ST theory in Ref.~\cite{2016CQGra..33m5002G}.
As in Ref.~\cite{2016CQGra..33m5002G}, we use a phenomenological
hybrid equation of state (EOS)
$P=P_{\rm c}+P_{\rm th}$, $\epsilon=\epsilon_{\rm c}
+\epsilon_{\rm th}$ with the {\em cold} part
\begin{eqnarray}
  &&\rho \le \rho_{\rm nuc}:~P_{\rm c}=K_1\rho^{\Gamma_1}
    \,,~~
    \epsilon_{\rm c}=\frac{K_1}{\Gamma_1-1}\rho^{\Gamma_1-1}\nonumber
    \\
  &&\rho > \rho_{\rm nuc}:~P_{\rm c}=K_2\rho^{\Gamma_2}\,,~~
    \epsilon_{\rm c}=\frac{K_2}{\Gamma_2-1}\rho^{\Gamma_2-1}+E_3
    \,,
\end{eqnarray}
where $\rho_{\rm nuc}=2\times 10^{14}\,{\rm g}\,{\rm cm}^{-3}$,
$K_1=4.9345\times 10^{14}\,{\rm [cgs]}$, $K_2$ and $E_3$ follow
from continuity; $\epsilon_{\rm th}$
measures the departure of the evolved internal energy
$\epsilon$ from the cold contribution and generates a thermal
pressure component
$P_{\rm th}=(\Gamma_{\rm th}-1)\rho \epsilon_{\rm th}$\,.
We thus have three parameters to specify the EOS.
As in Ref.~\cite{2016CQGra..33m5002G}, 
we consider $\Gamma_1 = \{1.28, 1.3, 1.32\}$ for
the subnuclear, $\Gamma_2 = \{2.5, 3\}$ for the supernuclear EOS
and $\Gamma_{\rm th} = \{1.35, 1.5\}$ for the thermal part
describing a mixture of relativistic and nonrelativistic gases.
For the conformal factor, we use the quadratic
Taylor expansion
commonly employed in the literature \cite{Damour:1993hw,Damour:1996ke}
and the potential endows the scalar field with a mass
$\mu$,
\begin{equation}
  F=\exp(-2\alpha_0\varphi-\beta_0\varphi^2)\,,~~~~~
  V=\hbar^{-2}\mu^2\,\varphi^2/2\,.
\end{equation}
The discretization, grid and boundary treatment are identical to those described in detail in Sec.~3 of Ref.~\cite{2016CQGra..33m5002G}.

\medskip
\noindent{\bf \em Simulations~--~}
%
%
For the simulations reported here,
we employ a uniform grid with $\Delta r=166\,{\rm m}$ inside
$r=40\,{\rm km}$ and logarithmically increasing grid spacing up to the
outer boundary at $9\times 10^5\,{\rm km}$.
As detailed in the Supplemental Material,
we observe convergence between first and second order, in agreement
with the use of first- and second-order accurate discretization techniques in
the code, resulting in a numerical uncertainy of about $4\,\%$ in
the wave signals reported below.

\begin{figure}[t]
  \includegraphics[height=165pt,trim={0cm 0.1cm 0cm 0cm},clip=true]{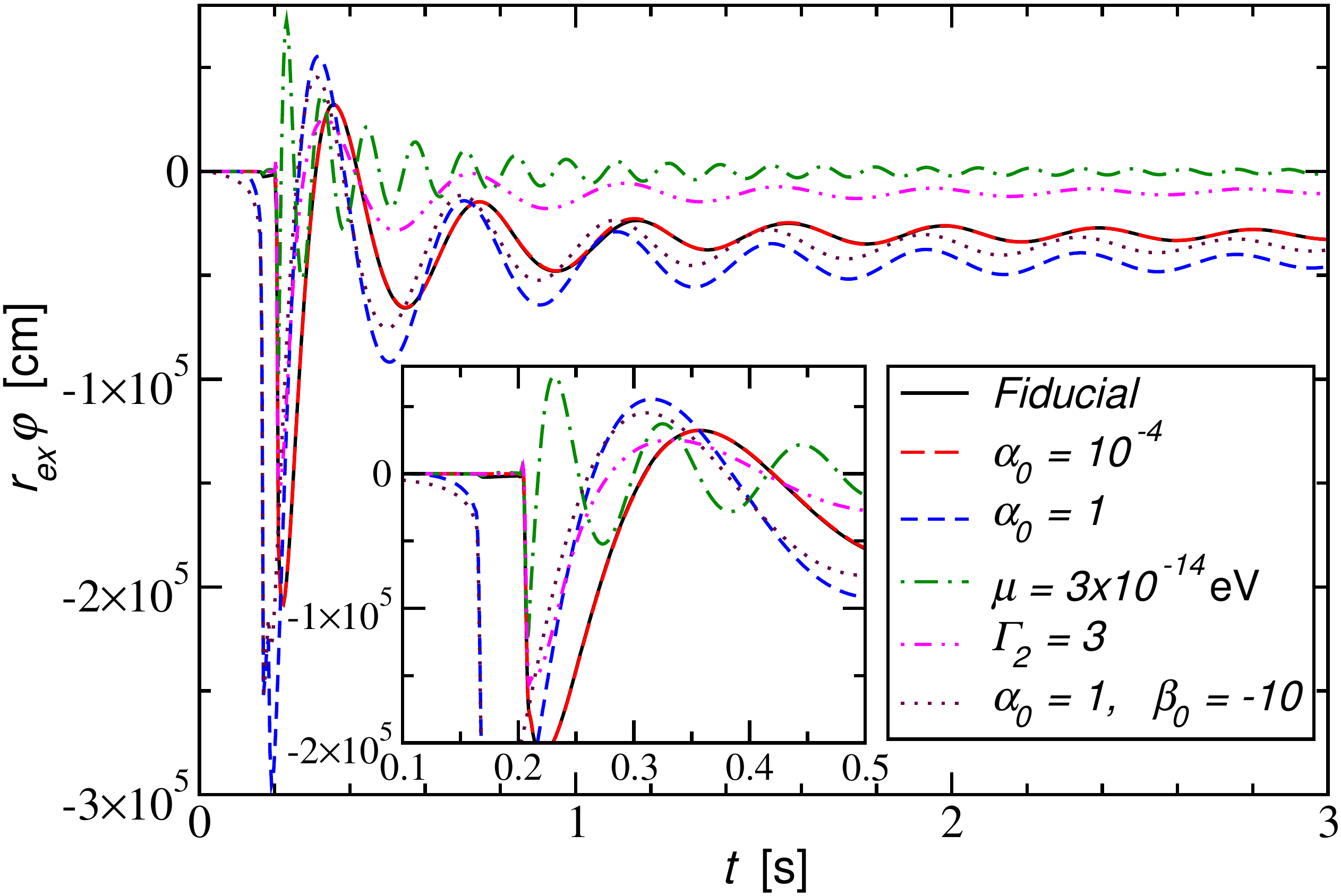}
  \caption{Waveforms extracted at $5\times 10^4\,{\rm km}$.
           The legend lists deviations from the fiducial parameters
           $\mu=10^{-14}\,{\rm eV},~\alpha_0=10^{-2},~\beta_0=-20,~
           \Gamma_1=1.3,~\Gamma_2=2.5,~\Gamma_{\rm th}=1.35$.}
  \label{fig:waveforms}
\end{figure}

All simulations start with the WH12 model
of the catalog of realistic
pre-SN models \cite{Woosley:2007as} with initially
vanishing scalar field.
The evolution is then characterized by
six parameters: the above-mentioned EOS
parameters $\Gamma_1$, $\Gamma_2$ and $\Gamma_{\rm th}$ as well as
mass $\mu$ of the scalar field and $\alpha_0,~\beta_0$ in the conformal function which we vary in the ranges
  $0\le\!\mu\!\le 10^{-13}\,{\rm eV}\,,~
  10^{-4}\le\!\alpha_0\!\le 1\,,~\text{and}~
  -25\le\!\beta_0\!\le -5\,.$
Our observations in these simulations are summarized as follows.
(i) The collapse dynamics are similar to the scenario displayed in the left panels in Fig.~4 in Ref.~\cite{2016CQGra..33m5002G}. As conjectured therein, the baryonic matter strongly affects the scalar radiation but itself is less sensitive to the scalar field.
(ii) For sufficiently negative $\beta_0$ the scalar field reaches amplitudes of the order of unity, independent of the EOS. Even in the massless case $\mu=0$, we observe this strong scalarization; the key impact of the massive field therefore lies in the weaker constraints on
$\alpha_0,\,\beta_0$ rather than a direct effect of terms involving $\mu$.
For illustration, we plot in Fig.~\ref{fig:waveforms} the wave signal
$r\varphi$ extracted at $5\times 10^4\,{\rm km}$ for various
parameter combinations.
These waveforms are to be compared with those obtained for
present observational bounds in the core collapse in massless ST
theory as shown in Fig.~6 of Ref.~\cite{2016CQGra..33m5002G}. The amplitudes
observed here are larger by $\sim 10^4$ for neutron star formation
from less massive progenitors and even exceed the strong
signals in black hole formation from more massive progenitors by $\sim 100$.
This \emph{hyperscalarization} of the collapsing stars
in massive ST theory (as compared with the more strongly
constrained massless case) and the
resulting substantially
larger GW signals are one of the key
results of this work. Translating this increase into improved
observational signatures for GW detectors, however, requires the
careful consideration of the signal's dispersion as it
propagates from the source to the detector; this is the
subject of the remainder of this Letter.

\medskip
%
%
\noindent{\bf \em Wave extraction and propagation~--~}
%
%
At large distances from the source, the dynamics of the scalar field are well approximated by the flat-space equation
  \mbox{$\partial_t^2 \varphi - \nabla^2 \varphi +
  \hbar^{-2}\mu^2\varphi =0\,,$}
which, in spherical symmetry,
reduces to a 1D wave equation for $\sigma\equiv r\varphi$.
Plane-wave solutions propagate with phase and group velocities ${v_{\textrm{g}/\textrm{p}}\!=\![1\!-\!(\omega_{*}^{2}/\omega^{2})]^{\pm 1/2}}$ for angular frequencies above ${\omega_{*}\!\equiv\!\mu /\hbar}$ but are exponentially damped for lower frequencies.

In the massless case (${\mu\!=\!0}$), the general solution for
$\sigma$ is the sum of an ingoing and an outgoing pulse propagating
at the speed of light. This makes interpreting the output of core
collapse simulations particularly simple; one extracts the scalar
field ${\sigma(t;r_{\textrm{ex}})}$ at a sufficiently large
\emph{extraction radius} $r_{\rm ex}$
and after imposing outgoing boundary
conditions the signal at ${r\!>\!r_{\textrm{ex}}}$ is
${\sigma(t;r)\!=\!\sigma(t\!-\!(r\!-\!r_{\textrm{ex}});r_{\textrm{ex}})}$.

In the massive case, the situation is complicated by the dispersive nature of wave propagation. However, an analytic solution for the field at large radii can still be written down, albeit in the frequency domain; ${\tilde{\sigma}(\omega;r)\!\equiv\!\int\textrm{d}t\,\sigma(t;r)\e^{\rmi\omega t}}$. 
The boundary conditions need to be modified for the massive case; frequencies ${|\omega|\!>\!\omega_{*}}$ propagate and we continue to impose the outgoing condition for these; however, frequencies ${|\omega|\!<\!\omega_{*}}$ are exponential (growing or damped) and we impose that these modes decay with radius.
These conditions determine the Fourier transform of the signal at large radii in terms of the signal on the extraction sphere (note the $\omega$ ranges),
\begin{equation}\label{eq:AnalyticSolution}
  \tilde{\sigma}(\omega;r) \!=\! \tilde{\sigma}(\omega;r_{\textrm{ex}}) \! \begin{cases} \!\e^{-\rmi \sqrt{\omega^{2}\textrm{--}\omega_{*}^{2}} (r-r_{\textrm{ex}}) } \!\!\!& \textrm{for}\, \omega\!<\!-\omega_{*} \\ \!\e^{+\rmi \sqrt{\omega^{2}\textrm{--}\omega_{*}^{2}} (r-r_{\textrm{ex}}) } \!\!& \textrm{for}\, \omega\!>\!-\omega_{*}\,. \end{cases}
\end{equation}
Note that the power spectrum $|\tilde{\sigma}(\omega;r)|^{2}$ is unchanged during propagation except for the exponential suppression of frequencies ${ \left| \omega  \right| \!<\!\omega_{*}}$.

\begin{figure*} 
\centering
\includegraphics[width=1.0\textwidth, trim={0.3cm 1.3cm 1.0cm 0.0cm}]{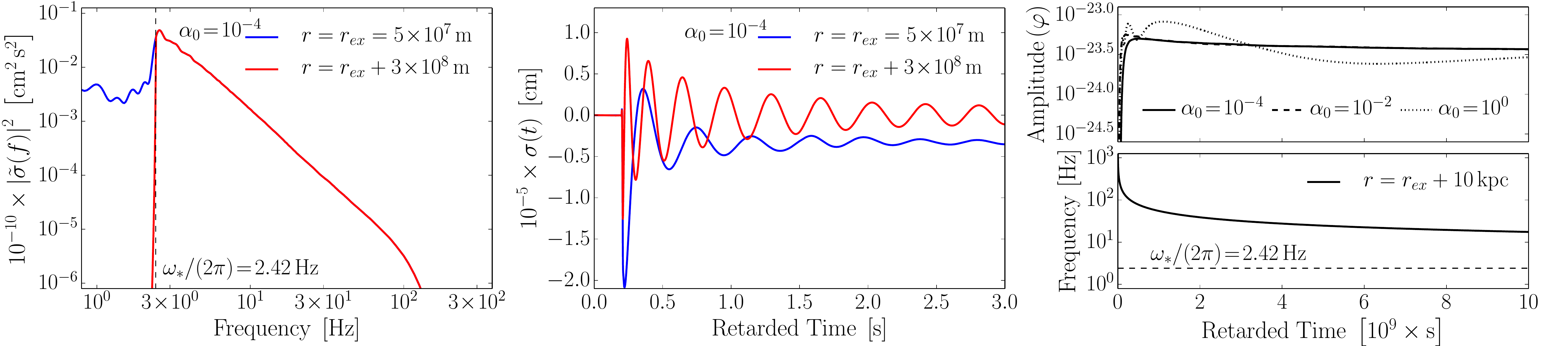}
\caption{\emph{Left panel:} The frequency-domain power spectrum of the scalar field $\sigma\!\equiv\!r\varphi$ at the extraction sphere and $1\,\textrm{light second}$ further out; the exponential decay of frequencies $f\!<\!\omega_{*}/(2\pi)$ can be clearly seen. This simulation was performed for a $12\,M_{\odot}$ star with $\mu\!=\!10^{-14}\,\textrm{eV}$, $\alpha_{0}\!=\!10^{-4}$, and $\beta_0\!=\!-20$. 
\emph{Center panel:} The time-domain scalar field profiles for the two curves shown in the left panel; during the $1\,\textrm{s}$ of propagation the signal becomes increasingly oscillatory, and the long-lived memory effect is exponentially suppressed. \emph{Right panels:} The amplitude (top) and frequency (bottom) as functions of time for the scalar field $\varphi$ from the same simulation as the other panels but at a distance of $10\,\textrm{kpc}$ (it is not practical to plot the long, highly oscillatory time-domain signals at large distances). Also shown by the dotted and dashed curves are the amplitude profiles from other simulations using $\alpha_{0}\!=\!10^{-2}$ and $\alpha_{0}\!=\!10^{0}$; the amplitude of the scalar field depends relatively weakly on $\alpha_{0}$.
For the simulations shown here, the energy radiated in scalar GWs is
$\sim 10^{-3}\,M_{\odot}$.
\label{fig:scalar_profiles}}
\end{figure*}

As signals propagate, they spread out in time, but the frequency content above the critical frequency $\omega_{*}$ remains unchanged. Consequently, the number of wave cycles in the signal increases with propagation distance;
cf.~Fig.~\ref{fig:scalar_profiles}. In the limit of large distances (relevant for LIGO observations of galactic supernovae) the signals are highly oscillatory, i.e.\ the phase varies much more rapidly than the frequency, and the inverse Fourier transform of Eq.~\eqref{eq:AnalyticSolution} may be evaluated in the \emph{stationary phase approximation} (SPA \cite{Bender:1978:AMM}). At each instant, the signal is quasimonochromatic with frequency
\begin{equation}\label{eq:OmegaSPA}
  \Omega(t)=\omega_{*}/\sqrt{1-\left[(r-r_{\textrm{ex}})/t\right]^{2}}\quad\textrm{for}\; t>r\!-\!r_{\textrm{ex}}\,.
\end{equation}
This time-frequency structure sounds like an \emph{inverse chirp}, with high frequencies arriving before low ones. The origin of this structure can be understood by noting that each frequency component arrives after the travel time of the associated group velocity. Using the SPA the time domain signal is given by ${\sigma(t,r)\!=\!A(t,r) \cos \phi(t,r)}$, where
\begin{align}
\phi(t,r) &= \sqrt{\Omega^{2}-\omega_{*}^{2}}(r-r_{\textrm{ex}})-\Omega t -\frac{\pi}{4}+\textrm{Arg}[\tilde{\sigma}(\Omega,r_{\textrm{ex}})] \,, \nonumber\\
A(t,r) &= \sqrt{\frac{2}{\pi}} \frac{(\Omega^{2}-\omega^{2}_{*})^{3/4}}{\omega_{*}(r-r_{\textrm{ex}})^{1/2}}\, \textrm{Abs}[\tilde{\sigma}(\Omega,r_{\textrm{ex}})]\,,\label{eq:SPA}
\end{align}
and the SPA frequency $\Omega(t)$ is given by Eq.~\eqref{eq:OmegaSPA}.

The Jordan frame metric perturbation is determined by the scalar field $\varphi$ (the tensorial GW degrees of freedom vanish in spherical symmetry). Any GW detector, small compared to the GW wavelength ${\lambda\!=\!2\pi/\omega}$, measures the \emph{electric} components of the Riemann tensor $R_{0i0j}$ \cite{Will:2014kxa}. In massless ST theory, this 3-tensor is transverse to the GW wave vector, ${R_{0i0j}\!\propto\!\delta_{ij}\!-\!k_{i}k_{j}}$, with strain amplitude ${h_{\textrm{B}}\!=\!2\alpha_{0}\varphi}$ (this is called a \emph{breathing} mode). In massive ST theory, there is an additional \emph{longitudinal} mode, ${R_{0i0j}\!\propto\!k_{i}k_{j}}$, with suppressed amplitude ${h_{\textrm{L}}\!=\!(\omega_{*}/\omega)^{2}h_{\textrm{B}}}$.
A GW interferometer responds identically (up to a sign) to both of these polarizations meaning they cannot be distinguished
\cite{Will:2014kxa}; henceforth we refer to the overall measurable \emph{scalar} signal with amplitude ${h_{\textrm{S}}\!=\!h_{\textrm{B}}\!-\!h_{\textrm{L}}\!=\!2\alpha_{0}[1\!-\!(\omega_{*}/\omega)^{2}]\varphi}$. In practice this factor reduces the strain only by at most a few percent at $t\lesssim 10^{10}\,{\rm s}$.

\medskip
%
%
\noindent{\bf \em LIGO observations~--~}
%
%
GW signals from stellar collapse in ST theory may show up in several ways in existing LIGO-Virgo searches.
In each case there is, in principle, a \emph{smoking gun} which allows the signal to be distinguished from other types of sources. 
Here, it is argued that a new dedicated program to search for ST core collapse signals is not needed; however, the results of this work should be kept in mind in analyzing results from existing searches.

\vspace{0.1cm}
\noindent{\em Monochromatic searches~--~} The highly dispersed signal [described by Eq.~\eqref{eq:SPA}, see right panels in Fig.~\ref{fig:scalar_profiles}] at large distances can last for many years and is nearly monochromatic on time scales of ${\lesssim 1\,\textrm{month}}$.
Quasimonochromatic GWs with slowly evolving
frequency may also be generated by rapidly rotating nonaxisymmetric neutron
stars; the scalar signals described in this Letter can be distinguished from
neutron stars by the scalar polarization content and the highly
characteristic frequency evolution described in Eq.~(\ref{eq:OmegaSPA}).

These signals may be detected by existing monochromatic searches
and allow for the determination of the scalar mass from the
frequency change $\dot{f}$.
The signals may show up in \emph{all-sky} searches; however, greater sensitivities can be achieved via \emph{directed} searches at known nearby supernovae (all-sky searches achieved sensitivities that constrain ${h\!\lesssim\!9.7\!\times\!10^{-25}}$\cite{Abbott:2016udd}, whereas model-based, directed searches at a supernova remnant have achieved sensitivities of ${h\!\lesssim\!2.3\!\times\!10^{-25}}$ \cite{Abbott:2017mwl} at frequencies $\sim\!150\,\textrm{Hz}$). Methods to detect signals of any polarization content have recently been presented in
Ref.~\cite{Isi:2017equ}; note that interferometers are a factor $\sim 2$ less sensitive to scalar than tensor GWs.
A directed search should begin within a few months to years of the supernova observation and may last for decades with sensitivity improving as $\textrm{time}^{-1/2}$ (see the amplitude as a function of time in Fig.~\ref{fig:scalar_profiles}). 
In fact, the amplitude can remain at detectable levels for so long that directed searches aimed at historical nearby supernovae (e.g.\ SN1987A\footnote{For $\mu=10^{-14}\,{\rm eV}$, for example, we obtain for SN1987A a frequency $\Omega/(2\pi)\approx 128\,{\rm Hz}$
and rate of change $\dot{\Omega}/(2\pi)\approx 2\,{\rm Hz/yr}$,
using distance $D:=r-r_{\rm ex}=51.2\,{\rm kpc}$ and time
$t-D=30\,{\rm yr}$.}) may be worthwhile; 
a nondetection from such a search can place the most stringent constraints to date on certain regions of the massive ST parameter space, $(\mu,\alpha_{0},\beta_{0})$.

In any monochromatic search there would be two smoking gun features indicating an origin of hyperscalarized core collapse in massive ST theory: the scalar polarization content, and the long signal duration with gradual frequency evolution according to Eq.~\eqref{eq:OmegaSPA}.
Our simulations suggest that the intrinsic amplitude of the scalar
field is insensitive to $\alpha_{0}$,
$\beta_0$, and $\mu$ over wide parameter ranges.
However, the GW strain scales linearly with the coupling; $h\!\propto\!\alpha_{0}\varphi$. Extrapolating the results in Fig.~\ref{fig:scalar_profiles} suggests that if a supernova at $10\,\textrm{kpc}$ were to be observed and followed up by a directed monochromatic search by aLIGO at design sensitivity, the coupling could be constrained to ${\alpha_{0}\!\lesssim\!3\!\times10^{-4}}$ (assuming no signal was in fact observed) which compares favorably with the impressive Cassini bound in the massless case \cite{2003Natur.425..374B}.

\vspace{0.1cm}
\noindent{\em Stochastic searches~--~} As shown above, stellar core collapse in massive ST theory can generate large amplitude signals, allowing them to be detected at greater distances. 
However, the signals propagate dispersively, spreading out in time and developing a sharp spectral cutoff at the frequency of the scalar mass.
The long duration signals from distant sources can overlap to form a stochastic background of scalar GWs with a characteristic spectral shape around this frequency. 
A detailed analysis of this stochastic signal
covering a wider range of ST parameters and progenitor models
will be presented in Ref.~\cite{INPREP}.

\vspace{0.1cm}
\noindent{\em Burst searches~--~} If the scalar field is light (${\mu\!\lesssim\!10^{-20}\textrm{eV}}$) then signals originating within the galaxy will not be significantly dispersed [e.g.\ the spread in arrival times across the LIGO bandwidth, ${(10\,\textrm{--}\,10^{3})\,\textrm{Hz}}$, for a source at ${10\,\textrm{kpc}}$ is ${\lesssim\!1\,\textrm{s}}$]. These short-duration, burstlike scalar GW signals may be detected using strategies similar to those used to search for standard core collapse supernovae in GR. However, for these light scalar fields the observational constraints on the coupling constants $\alpha_{0}$ and $\beta_{0}$ rule out the hyperscalarized signals shown in Fig.~\ref{fig:waveforms} and the amplitudes are similar to those reported in Ref.~\cite{2016CQGra..33m5002G}.

\medskip
%
%
\noindent{\bf \em Discussion~--~}
%
%
The main results of our work are the following points.
(i) Weaker constraints on the coupling parameters $\alpha_0$,
$\beta_0$ in ST theory with scalar masses $\mu \gtrsim
10^{-15}\,{\rm eV}$ allow for scalarization in stellar
core collapse orders of magnitude above what has been
found in massless ST theory. The scalar
signature is rather insensitive to the EOS parameters
and varies only weakly with the ST parameters $\alpha_0$
and $\beta_0$ for sufficiently negative $\beta_0$.
(ii) The strong scalar GW signal disperses as it
propagates over astrophysical distances, turning it into
an inverse chirp signal spread out over years with a near
monochromatic signature on time scales of $\sim1\,\text{month}$.
(iii) We identify three existing GW search strategies
(continuous wave, stochastic and burst searches)
that have the capacity to observe these signals
for galactic sources or infer unprecedented bounds
on the massive ST theory's parameter space through nondetection.

The dispersion
of the signal has two significant consequences.
(i) While the number of individually observable events may not change
significantly
from pure GR expectations (a few per century,
largely in the Milky Way and Magellanic Clouds),
each event remains visible for years or even centuries,
vastly increasing the number of sources visible {\em now}.
(ii) The signal
to be detected is largely insensitive to details
of the original source. Instead,
it is mainly characterized by the overall magnitude
of the scalarization and the ST parameters, most notably the mass $\mu$.
We tentatively conjecture that other prominent astrophysical
sources, such as a NS binary inspiral and merger, may result in a similar
inverse-chirp imprint on the GW signal in massive ST theory.
A natural extension of our work is the exploration of other
theories of gravity with massive degrees of freedom
(e.g.~\cite{Ramazanoglu:2017xbl}), 
but the results reported here already 
demonstrate the qualitatively new range of opportunities
offered in this regard by the dawn of GW astronomy.

%
%

\vspace{0.05cm}
\noindent{\bf \em Acknowledgments~--~}
This work was supported by the
H2020-ERC-2014-CoG Grant No.~646597,
STFC Consolidator Grant No.~ST/L000636/1,
NWO-Rubicon Grant No.~RG86688,
H2020-MSCA-RISE-2015 Grant No.~690904,
NSF-XSEDE Grant No.~PHY-090003,
NSF PHY-1151197,
NSF XSEDE allocation TG-PHY100033,
and DAMTP's Cosmos2 Computer system.
D.G. is supported by NASA through Einstein Postdoctoral Fellowship
Grant No. PF6-170152 by the Chandra X-ray Center, operated by the Smithsonian Astrophysical Observatory for NASA under
Contract No.~NAS8-03060.

%

\widetext
\begin{center}
  \textbf{\large Supplemental Material: Code Tests} \\
\end{center}

\setcounter{equation}{0}
\setcounter{figure}{0}
\setcounter{table}{0}
\setcounter{page}{1}
\makeatletter
\renewcommand{\theequation}{S\arabic{equation}}
\renewcommand{\thefigure}{S\arabic{figure}}
\renewcommand{\bibnumfmt}[1]{[S#1]}
\renewcommand{\citenumfont}[1]{S#1}

In order to test the the code for stellar collapse in massive
scalar-tensor (ST) theory of gravity, we have repeated the convergence
analysis displayed in Fig.~3 of \cite{sup2016CQGra..33m5002G} but now
using a massive scalar field with $\mu = 10^{-14}~{\rm eV}$
and $\alpha_0=10^{-4}$ and $\beta_0=-20$.
We observe the same convergence between first and second order,
in agreement with the first and second order schemes used in
the code.

\begin{figure}[h]
  \vspace{1cm}
  \includegraphics[height=165pt,trim={0cm 0.1cm 0cm 0cm},clip=true]{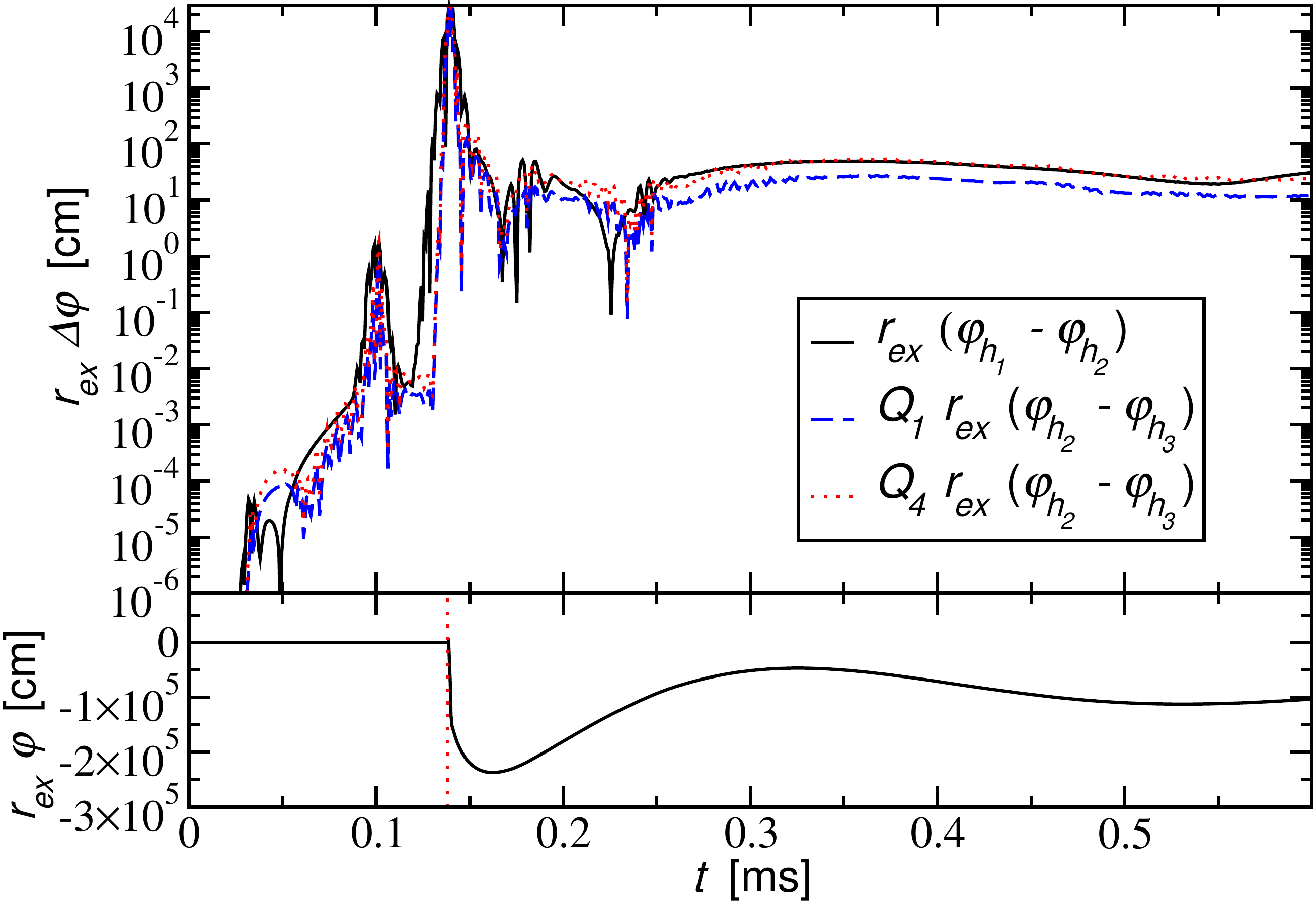}
  \caption{Convergence of the wave signal at $r_{\rm ex}=3\times 10^4~{\rm km}$ from a typical, strongly scalarized collapse of the WH12 profile with $\Gamma_1=1.3,~\Gamma_2=2.5,~\Gamma_{\rm th} =1.35,~\alpha_0=10^{-4},~\beta_0=-20$. The solid curve shows the difference of the coarse and medium resolution runs and is compared with that between medium and high resolution rescaled for first-order (dashed) and second-order (dotted curve) convergence factor. For reference, we show the signal $r_{\rm ex}\varphi$ in the bottom panel where the vertical dotted line at $t-r_{\rm ex}=38\,{\rm ms}$ marks the core bounce.}
  \label{fig:conv}
\end{figure}

As a further test, we have evolved the $12~M_{\odot}$
zero-age-main-sequence progenitor WH12 of the catalog of realistic
pre-SN models \cite{supWoosley:2007as}
for the same $\mu$, $\alpha_0$ and $\beta_0$,
employing a uniform grid with
$\Delta r$ inside $r=40\,{\rm km}$ and logarithmically increasing
grid spacing up to the outer boundary at $1.8\times 10^5\,{\rm km}$.
Convergence of $r\varphi$ extracted at $r_{\rm ex}=
3\times 10^9\,{\rm cm}$ is tested with three different resolutions $\Delta r_1=250\,{\rm m}$, $\Delta r_2=125\,{\rm m}$, $\Delta r=62.5\,{\rm m}$
in the interior
and a total number of $N_1=5\,000$,
$N_2=10\,000$, $N_3=20\,000$ grid points, respectively, so
that the differences between high, medium and low resolution
are expected to scale with
$Q_1=2$ for first and $Q_2=4$ for second-order convergence.
This expectation is borne out by Fig.~\ref{fig:conv} where we study
the convergence of the strong peak signal generated at core
bounce at $t-r_{\rm ex}\approx 38~{\rm ms}$ which dominates
all our wave signals. The good agreement between the solid and
dotted curves demonstrates convergence close to second order and
implies a discretization error of about $6~\%$ ($3~\%$) for
coarse (medium) resolution.
In the simulations used for our study,
we use $\Delta r=166~{\rm m}$ and extend the
outer grid to $9\times 10^5~{\rm km}$ while keeping the resolution in the extraction zone unchanged.

%

%

\end{document}